\documentclass[a4paper,11pt]{article}
\usepackage{jinstpub} 
\usepackage{makecell}
\usepackage{graphicx}
\usepackage{subcaption}
\usepackage{afterpage}

\usepackage{float}

\usepackage{booktabs}

\title{\boldmath Comparative Performance Analysis of  Crystals in Total-Body PET Scanners: Monte-Carlo Simulation Study with Different Materials and Geometry}

\author[1]{D. Choudhary\note{Corresponding author.}}
\author{and S. Nag}
\affiliation{Department of Physics, Indian Institute of Technology (Banaras Hindu University), Varanasi, India\\}

\emailAdd{deepa.student.phy20@itbhu.ac.in}

\abstract{Total-Body PET (TB-PET) scanners mark a significant leap in medical diagnostics, especially uEXPLORER, the world's first TB-PET, spanning 194 cm axially, showing remarkable sensitivity and spatial resolution. This study uses Monte Carlo simulation toolkit Geant4 to assess detector crystals of various configurations and materials. We focus on critical parameters—sensitivity, intrinsic coincident time resolution (CTR), and energy resolution—across three crystal designs: 1) standard LYSO Crystals, the baseline. 2) 0.1\% Mg, 1\% Ce Doped \(Gd_3Al_2Ga_3O_{12}\) (Mg,Ce:GAGG), an alternative material. 3) LYSO pyramid-shaped crystals, a geometric variant same dimensions. The study is based on the geometric configuration of the uEXPLORER. In our study, pyramid-shaped LYSO crystals demonstrated superior performance with an impressive CTR of 42 ps. PET detectors using doped GAGG crystals showed a 6\% poorer intrinsic CTR as compared to LYSO. However, Mg,Ce:GAGG outperformed LYSO in energy resolution by 25\%, while LYSO cuboidal crystals achieved approximately 37\% better sensitivity than Mg,Ce:GAGG crystals. Our findings illustrate how various crystal materials and geometries affect PET scanner performance, highlighting the trade-offs between sensitivity, coincidence time resolution (CTR), and energy resolution. These insights can serve as important factors for the design of future total-body PET (TB-PET) systems}


\begin{document}
\maketitle
\flushbottom

\section{Introduction}
\label{sec:intro}

\subsection{Positron Emission Tomography}
Positron Emission Tomography (PET) is a widely recognized non-invasive imaging technique that has seen extraordinary progress in recent years, particularly with the advent of TB-PET scanners like uEXPLORER. These advanced systems have shown significant improvements in sensitivity, signal-to-noise ratio, and lesion detectability compared to conventional PET systems \cite{ref1,cherry2018total,humanstudy,Lan2021}. To optimize the performance of PET scanners, focusing on the development of scintillation crystals and their geometries is essential. These crystals play a critical role in PET imaging by converting gamma rays into optical photons, which are then detected to produce the final image.

The performance of a PET scanner also largely depends on the detector crystals being used \cite{Yu2022,Ghabrial2018}. PET scanners continuously aim to improve Coincidence Time Resolution (CTR) for enhanced Time-of-Flight (TOF) applications, as well as sensitivity and energy resolution for more accurate gamma energy detection. To achieve these improvements, we focus on optimizing crystal performance by exploring different factors such as shape and material composition. Detector crystal's configuration has the ability to influence the energy resolution \cite{energyreso,ElHamli2022} as well as the time resolution \cite{auffrey,danevich2014impact}. The variation of these parameters owing to change in detector configuration can be studied by Monte-Carlo Simulations. We have employed Geant4 in this particular study \cite{ref2}. Geant4, a state-of-the-art simulation toolkit, has emerged as a powerful tool for modeling complex physical processes in radiation detection and medical imaging. It gives us the flexibility to simulate an accurate model of a PET scanner and modify different scintillation properties as well as complex detector geometry. We could use the existing data on standard materials to extend our simulations for different configurations. In this study, we have used the geometry of uEXPLORER TB-PET as the base model to evaluate sensitivity for different detector crystals. Its design parameters are given in table \ref{tab:2}. There have been studies regarding validating uEXPLORER parameters like \cite{Rezaei2023,gate1}. We have used Geant4 v11.2.1 instead of Gate in our case. The background for PET simulation in Geant4 is very well provided by Abdella M Ahmed et al. \cite{Ahmed2020}.

\subsection{Scintillator Configurations}
\begin{figure}[htbp]
\centering
\includegraphics[width=.4\textwidth]{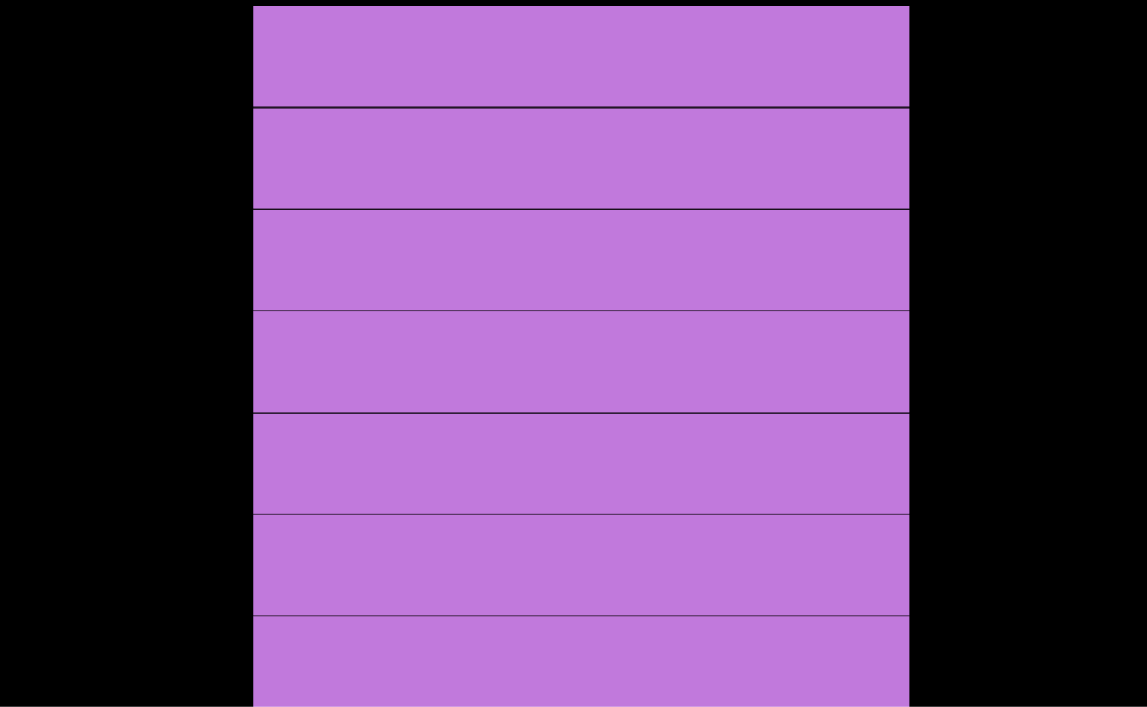}
\qquad
\includegraphics[width=.4\textwidth]{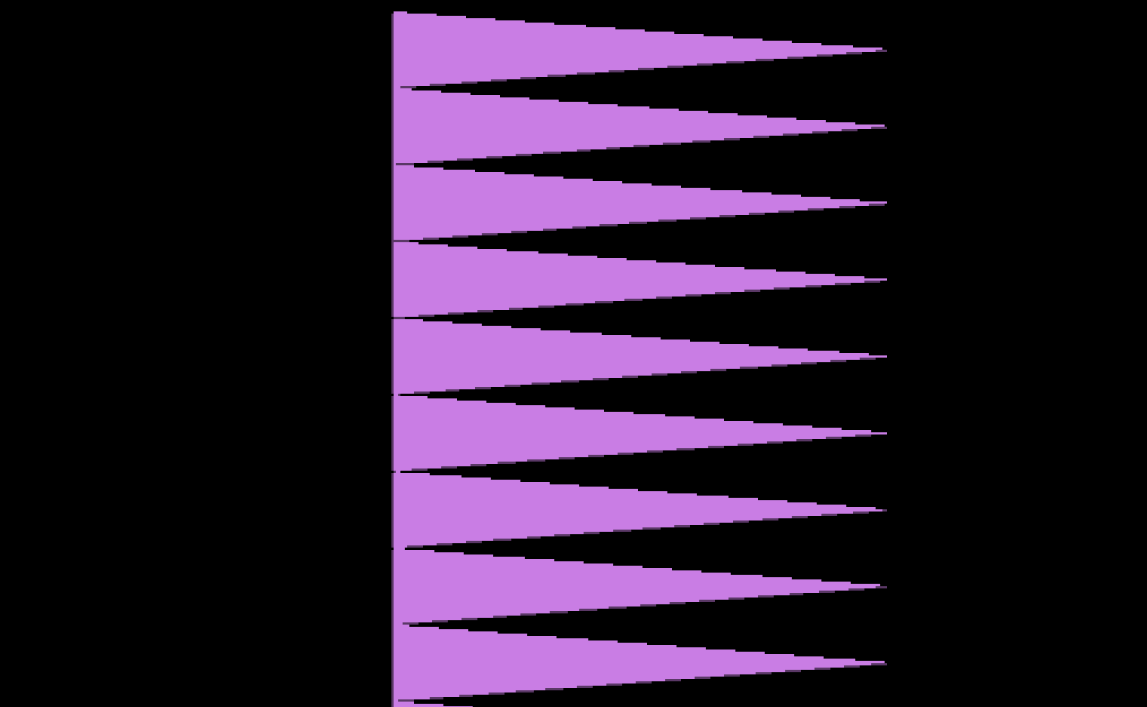}
\caption{Magnified side view of cuboid and pyramid-shaped detector crystals ($2.76\times 2.76\times 18.1 $mm) simulated in Geant4. \label{fig:1}}
\end{figure}

For all the available crystals, LYSO has gained popularity for its high density, good light yield, and faster scintillation decay time \cite{lyso,lyso1}. There are still concerns regarding its intrinsic radioactivity, high cost, and potential time and energy resolution improvement. Therefore, ongoing research explores alternative materials and geometries to enhance scanner performance \cite{Ozsahin2020,Ghabrial2018}. Among these, the multi-component garnet \(Gd_3Al_2Ga_3O_{12}\) doped with Cerium and Magnesium (GAGG) is one such promising material \cite{ref3}. While Mg co-doping does not enhance its light yield, it expedites the scintillation time response of the scintillator, which is crucial in TOF-related studies of the PET scanner \cite{KAMADA2017407,Dantelle2019,YOSHINO2017420}. We have used Mg,Ce:GAGG for its high light yield, faster decay decay time, and absence of internal radioactivity, which could make it a better alternative to LYSO. Moreover, Mg,Ce:GAGG also shows superior radiation hardness as compared to LYSO \cite{radhard}. Properties of both the materials are given in table \ref{tab:1}. The detector crystal's shape also influences the detector's timing property. The scintillating photons reflected off the pyramid surface will alter their trajectory and directly reach the coupling surface, enhancing light output and improving timing resolution. Cuboid-shaped crystals often require multiple photon reflections for absorption, potentially impacting timing resolution. To mitigate this, we explored pyramid-shaped crystals \cite{ref4}. While pyramid shapes can enhance light collection, the variations in the crystal’s cross-sectional area along its height can introduce position-dependent signal variations, which are needed to be considered. However detailed characterization of this effect in pyramid-shaped crystals is still limited in the literature. The simulated images of both of the crystal shapes is shown in figure \ref{fig:1}.

Some studies have been conducted with GAGG as a detector crystal in PET, like \cite{cherrygagg,Schneider2015,Lee2020}. In our research, we will consider the Ce, Mg-doped version of GAGG, which has a faster decay constant . We will also calculate the sensitivity, CTR, and energy resolution of a TB-PET made from pyramid-shaped LYSO and Mg,Ce:GAGG crystals.

\section{Materials and Methods}

\subsection{Simulation Setup}

\begin{figure}[H]
\centering
\includegraphics[width=0.56\textwidth]{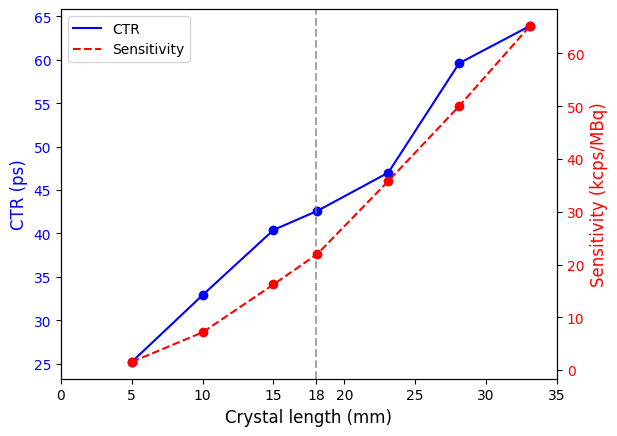}
\caption{Sensitivity and intrinsic CTR values for pyramidal crystals of different lengths. \label{fig2.b}}
\end{figure}

\begin{figure}[htbp]
\centering
\includegraphics[width=0.56\textwidth]{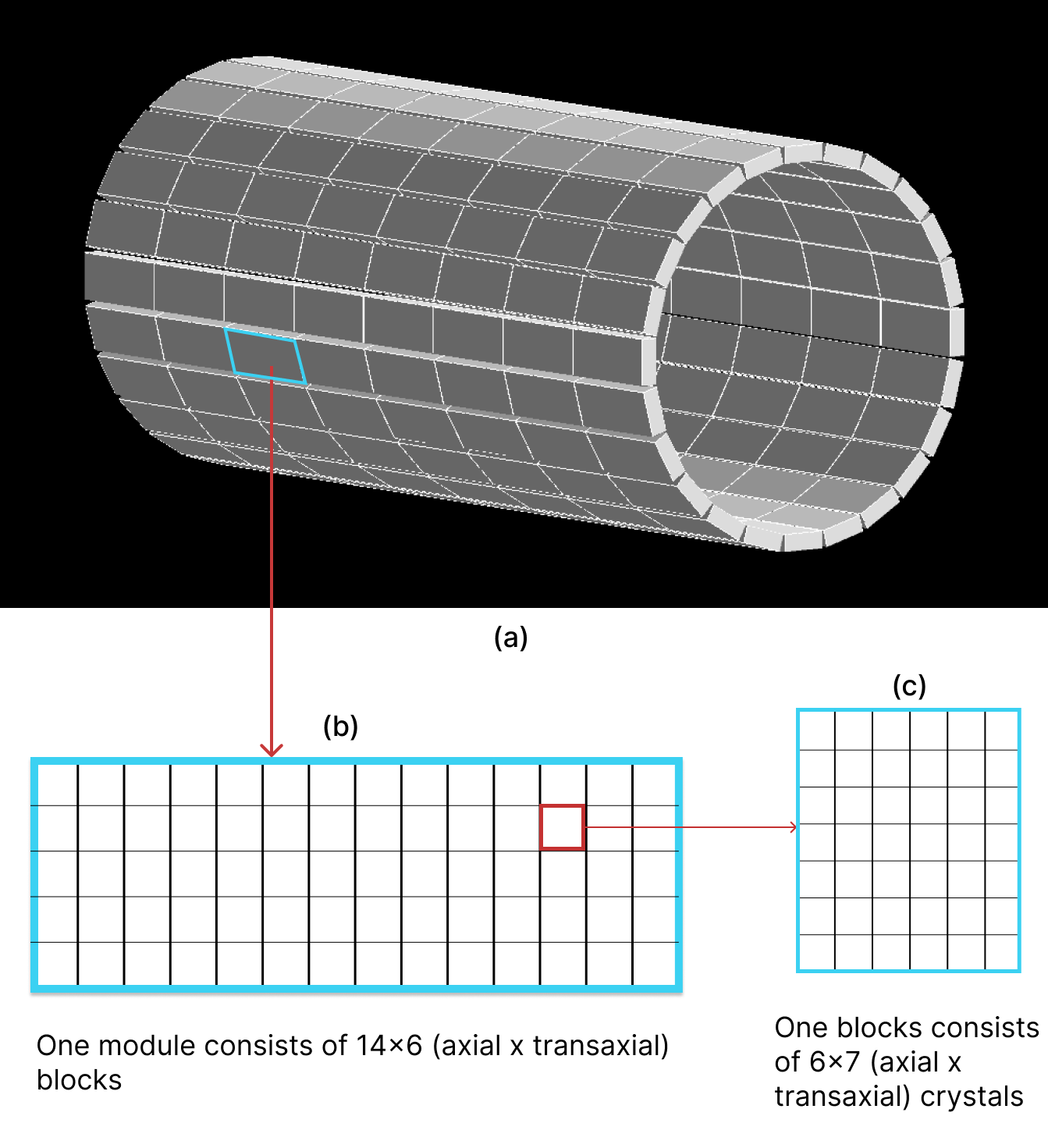}
\caption{a) Geant4 Simulation of axial view of EXPLORER PET b) module configuration c) block configuration.\label{fig2}}
\end{figure}

For the sensitivity test, we will be constructing three PET scanners with geometrical parameters
of uEXPLORER. Those are a) Actual uEXPLORER consisting LYSO crystals of length 18mm.
b) 0.1\% Mg, 1\% Ce Doped GAGG crystals of length 18mm. c) Pyramid-shaped LYSO crystal of length 18-mm. The pyramidal geometry of the detector is a trade-off between the CTR and sensitivity. With increasing crystal length, sensitivity improves but CTR deteriorates \cite{timingchoong2009, sensitivitymacdonald2013effects}; therefore, 18-20 mm crystal length is widely used for a better tradeoff between these two parameters. We study this correlation in pyramidal crystals as shown in figure \ref{fig2.b}. The geometrical parameters of the PET scanner are given in table \ref{tab:2}, and its visual representation is given in figure \ref{fig2}. The LYSO crystal has a density of 7.11 g/cm\textsuperscript{3}, and its composition includes 71.447\% of Lutetium, 4.034\% of Yttrium, 6.371\% of Silicon, and 18.148\% of Oxygen \cite{Rezaei2023}. The pyramid crystals were simulated for 18-mm length. 

\begin{table}[htbp]
\centering
\caption{Properties of LYSO and Mg,Ce:GAGG crystals \cite{luxim, Dantelle2019}\label{tab:1}}
\smallskip
\begin{tabular}{l c c}
\toprule
   {Properties} & {LYSO} & {0.1\% Mg, 1\% Ce doped GAGG} \\
\midrule
Density & 7.4 g/cm\textsuperscript{3} & 6.63 g/cm\textsuperscript{3} \\
Light Yield (Photons/MeV) & 32,000 & 44,000 \\
Decay time (in ns) & 41 & \makecell{45 (58\%) - Fast decay time \\ 135 (52\%) - Slow decay time} \\
\bottomrule
\end{tabular}
\end{table}

\begin{table}[htbp]
\centering
\caption{uEXPLORER PET design parameters.\label{tab:2}}
\smallskip
\begin{tabular}{c|c}
\toprule
Ring diameter	& 78.6 cm\\
Axial FOV & 194 cm\\
Crystal size	& $2.76\times 2.76\times 18.1 $mm\\
Crystal pitch&	$2.85\times 2.85 $mm\\
Crystals per block&	 $6 \times7$=42\\
Blocks per module&	 $14 \times5$=70\\
Number of rings&	 8\\
Number of modules per ring&	 24\\
Energy window&	 400-600 keV\\
Total number of crystals&	 564,480\\
\bottomrule
\end{tabular}
\end{table}

The Geant4 simulation toolkit was employed to mimic the actual performance of the scanner. Physics list G4EmStandardPhysics\_option3, G4DecayPhysics, and G4RadioactiveDecayPhysics were used to simulate the underlying physics involved. The optical physics module was disabled for this setup to save memory and system load. Then, a blurring module was used which assigned energy resolution to the simulated crystals. This module introduced Gaussian energy blur to the absorbed energy. It allowed us to achieve an energy resolution matching the experimental energy resolution. So, mean energy resolution value of 11.7\% \cite{Rezaei2023} for LYSO and 8.6\% in GAGG \cite{ref5,YOSHINO2017420} in reference to 511 keV was adopted, whereas a detection efficiency factor of 0.883 was added to the recorded pulses. Next, we conducted NEMA performance tests, the standard method for assessing PET system performance and comparing it across different systems. In this work, the NEMA-NU2-2018 standards \cite{nema} are used to first validate the sensitivity for uEXPLORER and subsequently evaluate the performance of other scanners on the same geometry.

For evaluating the timing and energy performance of different detector configurations, we used individual crystal(s) instead of the whole scanner to reduce the system load while generating optical photons. The details on the setup and evaluation of energy resolution and CTR are given in the section \ref{subsecCTR} and section \ref{subsecER}. All the simulations and post-analysis were done on Ubuntu 22.04.5, 12th Gen Intel Core i7-12700H × 20 processor.

\subsection{NEMA Performance Evaluation}
\subsubsection{Sensitivity}
This study investigated the sensitivity of a PET scanner to F-18 (fluorine-18) detection under increasing aluminum attenuation, following the NEMA-NU2 2018 standards \cite{nema}. A virtual cylindrical water phantom with dimensions of 700 mm length, 2 mm inner diameter, and 3.2 mm outer diameter was filled with a  F-18 and water mixture. Separate simulations progressively added aluminum layers of 2.5 mm thickness around the phantom, and the scanner's sensitivity for detecting F-18 gamma rays was calculated for each configuration. Sensitivity was assessed for two phantom positions: centered within the field of view (FOV) and radially offset by 10 cm. A simulated source activity of 3 MBq of F-18 was used throughout. The relationship between aluminum layer count (attenuation) and scanner sensitivity will be analyzed by plotting a curve and extrapolating to estimate the ideal sensitivity without attenuation. Additionally, a 1700 mm long and 2 mm thick polyethylene phantom, filled with 3 MBq of F-18 mixed with water and without an aluminum layer, was used to assess the scanner's sensitivity.
A total of  $2\times 10^7$ events was simulated, and list-mode coincidence data was evaluated to calculate the sensitivity using equation \ref{equation2}.

\begin{equation}
\text{Sensitivity} = \frac{\text{Total recorded counts}}{\text{Duration of scan }{\times \text{ Initial Activity}}}
\label{equation2}
\end{equation}

In Geant4, the duration of the scan is calculated by dividing the number of simulated events by the initial activity. So on simplification, sensitivity can now be calculated as:

\begin{equation}
\text{Sensitivity} = \frac{\text{Total recorded counts}}{\text{Total number of simulated events }}
\label{equation3}
\end{equation}

\subsection{Coincident Time Resolution (Intrinsic)}
\label{subsecCTR}
\begin{figure}[htbp]
\centering
\includegraphics[width=0.4\textwidth]{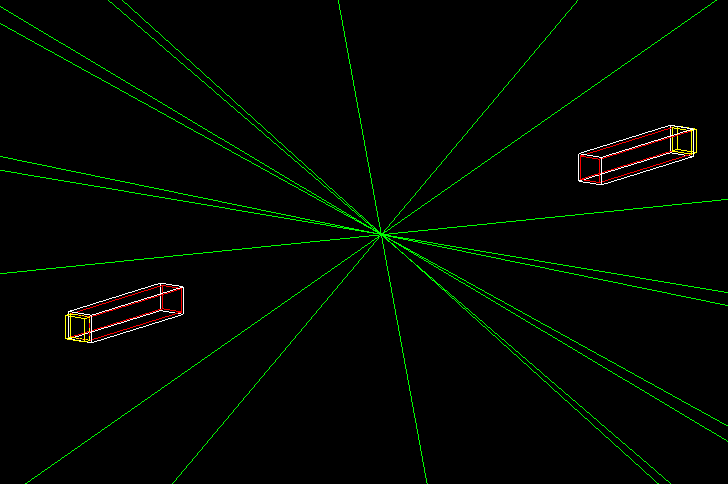}
\caption{Geant4 Simulation of CTR experimental setup.\label{fig:ctr}}
\end{figure}

For this study, we used two single crystals of the same length as crystals used in the PET scanner, i.e., $2.76\times 2.76\times 18.1 $mm, separated by a distance of 30 cm. Its visualization in geant4 is given in figure \ref{fig:ctr}. Each crystal was covered by an aluminum reflector, with surface type, finish, and model as dielectric\_metal, ground, and unified, respectively. Its reflectivity was set as 98-99\%, and a small air gap of 0.1 mm was introduced between the scintillator and the reflector surface. Optical Physics class was used to simulate the behavior of optical photons. One face of the scintillator was assigned as a 'sensitive detector' to capture the information of optical photons falling on it. A $^{22}$Na point source wrapped in a polyethylene cover was kept in the middle of the two crystals. The coincident 511 keV hits the opposite faces of the detector and generates optical photons. The time of hit was recorded by taking the first photon hit in each detector. The events were sorted for coincidence, and corresponding hit times were subtracted. The FWHM of those differences in hit times makes up the intrinsic coincident time resolution (CTR). No jitter or blurring was used in this case.

\subsection{Energy Resolution}
\label{subsecER}
We used a single scintillator of dimensions $2.76\times 2.76\times 18.1 $mm. An isotropic 511 keV gamma source was placed perpendicularly at a distance of 22 cm from the scintillator. This time, we focused more on using the SCINTILLATIONYIELD and RESOLUTIONSCALE properties of the scintillator. The data for light yield was taken from the literature \cite{luxim, YOSHINO2017420}. RESOLUTIONSCALE, determines the variance of the normal distribution of emitted scintillation photons for a given energy deposition in the scintillator \cite{opengate_optical_photons}. For LYSO and Mg,Ce:GAGG, the simulation will show 'better' energy resolution because we didn't account for electronic noise, temperature fluctuations, and other limiting factors. Therefore, we used energy resolution values for standard LYSO \cite{ref1} and Mg,Ce:GAGG crystals \cite{YOSHINO2017420, gagg} from the published data and calculate RESOLUTIONSCALE as per \cite{opengate_optical_photons} to match the experimental results. This cross-validation helped us to perform this evaluation on different shapes of the detector, without changing any scintillation property of the detector. The value of energy resolution is calculated using equation \ref{equation}
\begin{equation}
\text{Energy Resolution} = \frac{\text{FWHM of the photon distribution}}{\text{Mean number of photons}} \times 100\%
\label{equation}
\end{equation}

\section{Results}
\subsection{Sensitivity}
\begin{figure}[htbp]
    \centering
 \begin{minipage}{0.49\textwidth}
    \centering
    \includegraphics[width=1\textwidth]{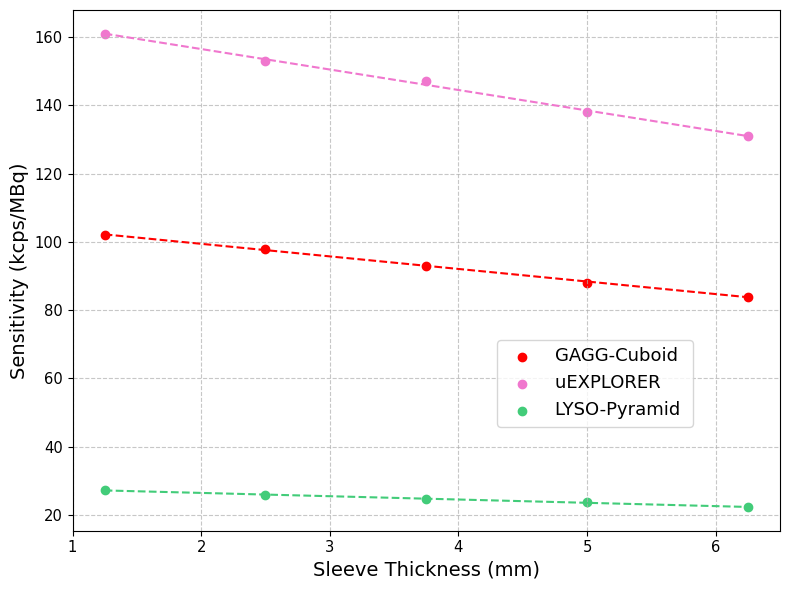}
    \\ \textbf{(a)}
    \end{minipage}
    \hfill
    \begin{minipage}{0.49\textwidth}
    \centering
    \includegraphics[width=1\textwidth]{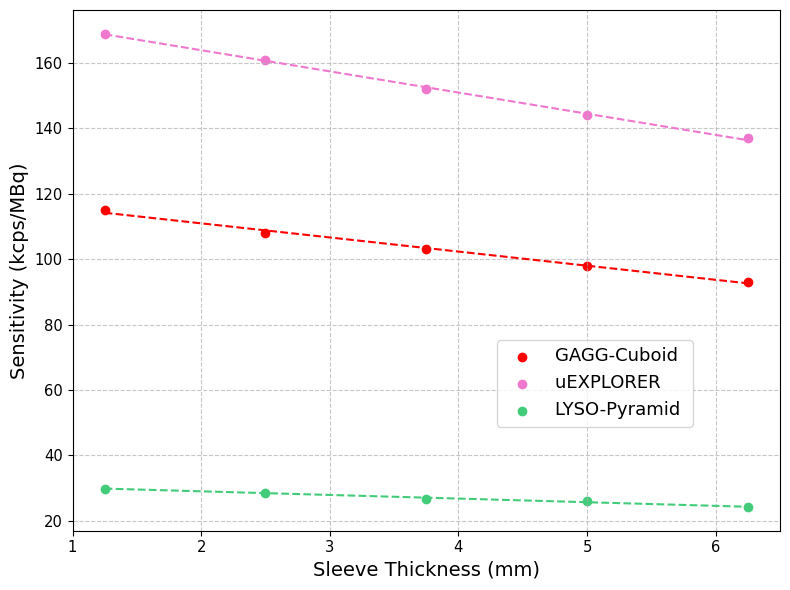}
    \\ \textbf{(b)}
    \end{minipage}
\caption{Sensitivity values when the source was at a) 10 cm offset b) at the center for 70 cm long phantom\label{fig:3}}
\end{figure}

\begin{figure}[htbp]
    \centering
 \begin{minipage}{0.49\textwidth}
    \centering
    \includegraphics[width=1\textwidth]{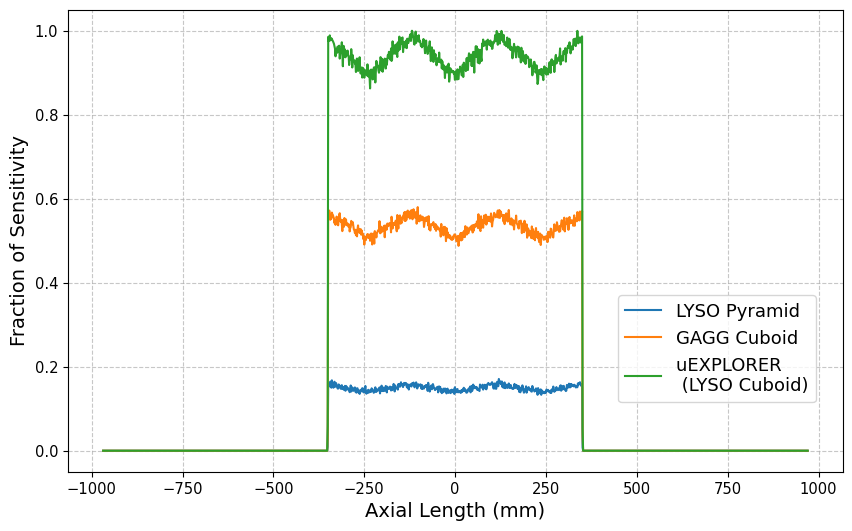}
    \\ \textbf{(a)}
    \end{minipage}
    \hfill
    \begin{minipage}{0.49\textwidth}
    \centering
    \includegraphics[width=1\textwidth]{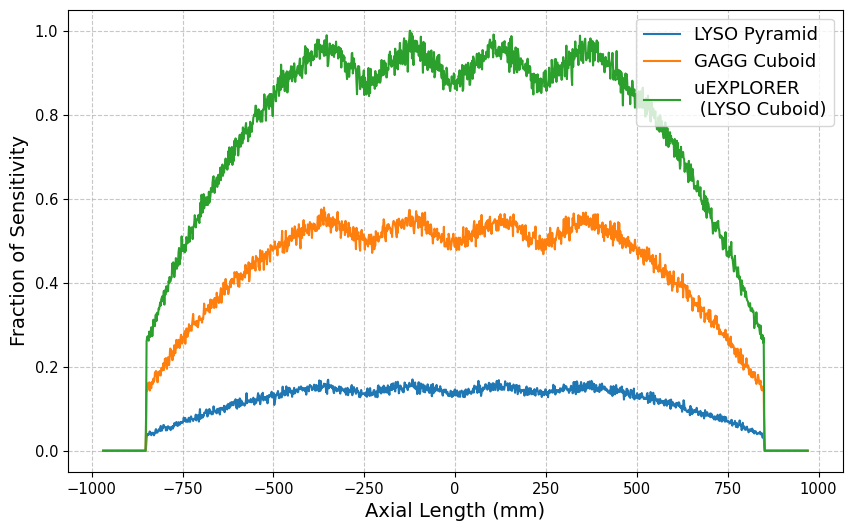}
    \\ \textbf{(b)}
    \end{minipage}
\caption{Axial sensitivity profiles of different PET scanners when the source was at a) 10 cm offset b) at the center\label{fig:4}}
\end{figure}

Table \ref{tab:i} presents the simulated sensitivity test results for 70 cm and 170 cm phantoms at two radial positions: the center of the FOV and 10 cm off-center. For uEXPLORER, our simulations gave results close to the experimental ones. The deviation was 1.13\% (0 cm offset)and 5.08\% (10 cm offset) for the 70 cm long phantom and 2.7\% (0 cm offset) and 2.1\% (10 cm offset) for the 170 cm long phantom. So, based on the same setup, we changed the crystal configuration to calculate the sensitivity for each of them. The sensitivity performance for each of the simulated scanners, with respect to sleeve thickness, is shown in the figure \ref{fig:3}. 
The simulated axial
sensitivity profiles for the 70 cm and 170 cm phantoms are presented in figure \ref{fig:4}. Mg,Ce:GAGG shows less sensitivity than LYSO because of lesser density and effective atomic number. Pyramid-shaped LYSO shows the least sensitivity of all due to lesser volume and the presence of dead space between the crystals. The reduction in sensitivity can be leveraged using Depth of Interaction (DOI) techniques, which we plan to incorporate in future work.

\begin{table}[htbp]
\centering
\caption{Total sensitivity for 70 cm and 170 cm line sources placed at the center and 10 cm off-center of the FOV across different simulated PET scanners\label{tab:i}}
\resizebox{\textwidth}{!}{%
    \begin{tabular}{lccc}
    \hline
      {PET Scanner} & {Transaxial Offset Position} & {Sensitivity- 70 cm phantom} & {Sensitivity- 170 cm phantom} \\
      \hline
      uEXPLORER (LYSO) & 0 cm & 176 kcps/MBq & 148 kcps/MBq \\
     
      & 10 cm & 168 kcps/MBq & 140 kcps/MBq \\
      \hline
      Mg,Ce:GAGG& 0 cm & 110 kcps/MBq & 77 kcps/MBq \\
     
      & 10 cm & 106 kcps/MBq & 74 kcps/MBq \\
      \hline
      Pyramidal-LYSO & 0 cm & 31 kcps/MBq & 28 kcps/MBq \\
     
      & 10 cm & 23 kcps/MBq & 22 kcps/MBq \\
      \hline
    \end{tabular}
}
\end{table}

\subsection{Intrinsic Coincident Time Resolution}

\begin{figure}[htbp]
\centering
\includegraphics[width=0.56\textwidth]{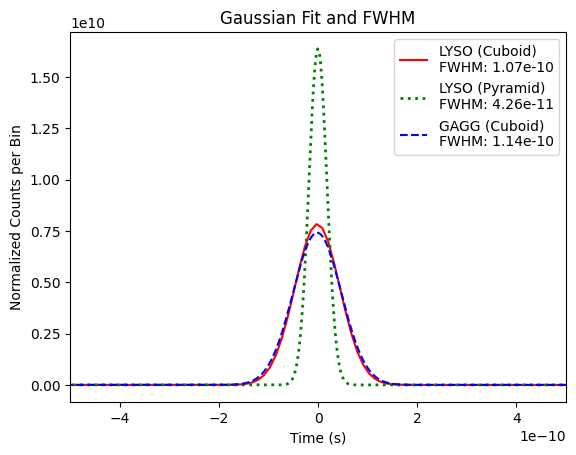}
\caption{ Intrinsic CTR values for different geometries of LYSO and GAGG crystals.
\label{fig:5}}
\end{figure}

The Full Width at Half Maximum (FWHM) of the time difference distribution of the first photon hits represents the intrinsic Coincidence Time Resolution (CTR) of our system. This intrinsic CTR is a core measure of the detector’s temporal performance, as it remains unaffected by external factors like electronic noise in the readout circuits, timing jitter in the photodetectors, variations in the crystal growth and surface treatment, temperature fluctuations, etc., Therefore, the CTR value we obtain is expected to be lower than what would be observed in real-world experimental conditions, as shown in figure \ref{fig:5}. 
The Mg,Ce:GAGG cuboidal crystal has an intrinsic CTR of 114 ps, compared to 107 ps for the LYSO cuboidal crystal. For the pyramid-shaped crystals, Mg,Ce:GAGG shows a very slight deterioration in CTR by 0.3 ps compared to LYSO. As this difference is minimal, the pyramid Mg,CeGAGG data was not shown separately in the plot. Overall, pyramid-shaped crystals exhibit a 60.6\% improvement in CTR as compared to cuboidal crystals.

Future work could involve incorporating real-world factors into the simulation to provide a more comprehensive comparison of CTR with experimental results.

\subsection{ Energy Resolution }

\begin{figure}[htbp]
    \centering
    \begin{subfigure}[b]{0.4\textwidth}
        \centering
        \includegraphics[width=\textwidth]{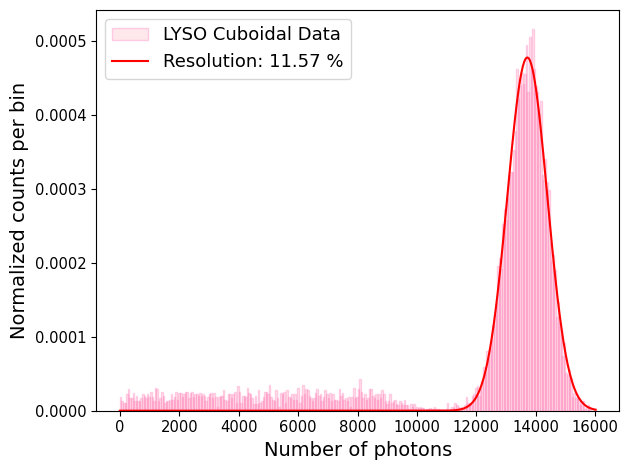}
        \caption{}
        \label{fig:figure1}
    \end{subfigure}
    \hspace{0.02\textwidth}
    \begin{subfigure}[b]{0.4\textwidth}
        \centering
        \includegraphics[width=\textwidth]{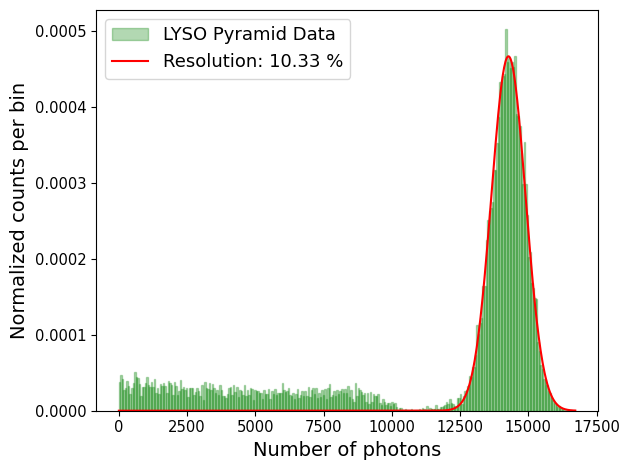}
      \caption{}
        \label{fig:figure2}
    \end{subfigure}
    \vskip\baselineskip
    \begin{subfigure}[b]{0.4\textwidth}
        \centering
        \includegraphics[width=\textwidth]{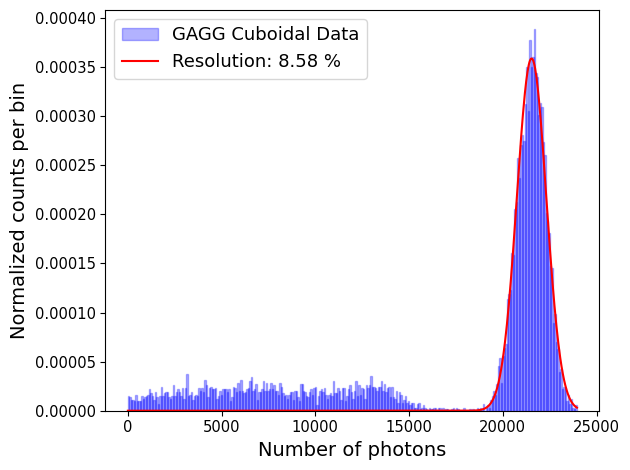}
       \caption{}
        \label{fig:figure3}
    \end{subfigure}
    \hspace{0.02\textwidth}
    \begin{subfigure}[b]{0.4\textwidth}
        \centering
        \includegraphics[width=\textwidth]{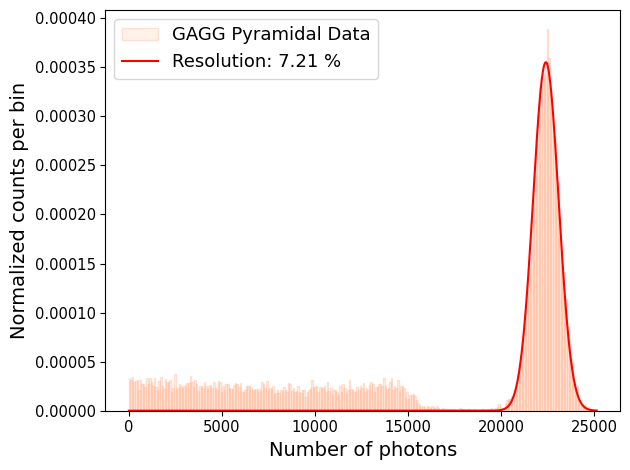}
        \caption{}
        \label{fig:figure4}
    \end{subfigure}
    \caption{Energy resolution plots for a) LYSO Cuboid, b) LYSO Pyramid, c) GAGG Cuboid, and d) GAGG Pyramid}
    \label{fig:fourfigures}
\end{figure}

In figure \ref{fig:fourfigures}, the x-axis represents the number of photons hitting the sensitive detector in each event. Mg,Ce:GAGG showed an energy resolution of 8.59\% for cuboidal crystals and 7.21\% for pyramid-shaped crystals, reflecting a 16.06\% improvement due to the change in shape. Similarly, LYSO had an energy resolution of 11.57\% for cuboidal crystals and 10.33\% for pyramid-shaped crystals, indicating a 10.71\% improvement with the shape modification. The better performance of pyramid crystals is due to better Light Collection Efficiency (LCE) in this configuration. This shape minimizes optical-photon reflections at the scintillator-reflector boundary, thereby increasing the likelihood of photon detection at the photocathode and enhancing energy resolution \cite{energyreso}. Overall, Mg,Ce:GAGG demonstrates much better energy resolution than the LYSO crystal.

\section{Conclusion}
In this work, we demonstrate that pyramid-shaped crystals of LYSO can improve intrinsic CTR by \textasciitilde 60.6\% and energy resolution by \textasciitilde 10.71\% compared to traditional cuboidal crystals, indicating their potential usage in TOF PET. Mg,Ce:GAGG demonstrated CTR of 114 ps, which is rather good (though not better) if compared with 107 ps of LYSO, outperforming many high-density scintillators like BGO. Mg,Ce:GAGG-cuboid also shows 25.7\% better energy resolution than LYSO-cuboid. Thus, The Mg,Ce-doped GAGG crystal emerges as a promising detector material due to its high light yield, finer energy resolution, and lack of internal radioactivity. Although both pyramid-shaped crystals and Mg,Ce:GAGG crystals exhibit reduced sensitivity, this limitation can be addressed using depth of interaction (DOI) techniques. In conclusion, both pyramid-shaped crystals and Mg, Ce doped GAGG crystals show great promise for PET applications and open the way for optimizing crystal configurations within PET scanners towards better image quality and quantitative accuracy in research.

\acknowledgments

The authors acknowledge Dr. Sushil Sharma,
Manish Das, Dr. Biswajit Das, Dr. S. Saha and Dr. Reimund Bayerlein for their fruitful
discussions. S. Nag acknowledges the financial
support from SERB-DST India under CRG
(File No.: CRG/2021/006671).




\bibliography{cite}

\end{document}